\documentclass[preprint,aps]{revtex4-1}
\pdfoutput=1

\usepackage{amsmath,amssymb,amsfonts,dcolumn,color,graphicx,graphics,latexsym,placeins,epsfig}
\usepackage{epsfig}
\usepackage{bm}
\usepackage{slashed}
\usepackage{latexsym}
\usepackage{natbib}
\usepackage{url}
\usepackage{dcolumn}
\usepackage{color}
\usepackage{amsfonts,amssymb,amsmath}
\usepackage{graphicx,epsfig}
\usepackage{psfrag}
\usepackage{subfigure}
\usepackage{tabularx}
\usepackage{hyperref}
\hypersetup{colorlinks=true}

\newcommand{\be}{\begin{equation}}
\newcommand{\ee}{\end{equation}}
\newcommand{\ba}{\begin{eqnarray}}
\newcommand{\ea}{\end{eqnarray}}

\begin{document}

\title{Constraints on ultralight axions from compact binary systems}
\author{Tanmay Kumar Poddar$^{1}$\footnote{tanmay@prl.res.in}}
\author{Subhendra Mohanty$^{1}$\footnote{mohanty@prl.res.in}}
\author{Soumya Jana$^{1,2}$\footnote{Soumya.Jana@etu.unige.ch}}
\affiliation{${}^{1}$ {\it Theoretical Physics Division, Physical Research Laboratory, Ahmedabad 380009, India} }

\affiliation{${}^{2}$ {\it D\'epartement de Physique Th\'eorique, Universit\'e de Gen\`eve, 24 quai Ernest Ansermet, 1211Gen\`eve 4, Switzerland} }


\begin{abstract}
Ultra light particles $(m_a \sim 10^{-21}eV-10^{-22}eV)$ with axion-like couplings to other particles can be candidates for fuzzy dark matter (FDM) if the axion decay constant $f_a\sim 10^{17}GeV$.  If a compact star  is immersed in such a low mass axionic potential it develops a long range field outside the star. This axionic  field is radiated away when the star is in a binary orbit.
The orbital period of a compact binary decays mainly due to the gravitational wave radiation, which was confirmed first in the Hulse-Taylor binary pulsar. The orbital period can also decay by radiation of other light particles like axions and axion like particles(ALPs).  For axionic radiation to take place,  the orbital frequency of the periodic motion of the binary system should be greater than the mass of the scalar particle which can be radiated. This implies that, for most of the observed binaries, particles with mass $m_a< 10^{-19}eV$ can be radiated, which includes FDM particles.  In this paper, we consider four compact binary systems: PSR J0348+0432, PSR J0737-3039, PSR J1738+0333, and PSR B1913+16 (Hulse Taylor Binary) and show that the observations of the decay in orbital period put the bound on axion decay constant, $f_a\lesssim \mathcal{O}(10^{11}GeV)$. This implies that Fuzzy Dark Matter cannot couple to gluons. 
 
\end{abstract}


\maketitle

\section{Introduction}
Axion was first introduced to solve the strong CP problem \cite{quinn,weinberg,wil,peccei}. The most stringent probe of the strong CP violation is the electric dipole moment of neutron. Quantum chromodynamics (QCD) is one of the possible theories which can explain the strong interaction. However, the theory has a problem known as the strong CP problem. We can write the QCD Lagrangian 
\begin{equation}
\mathcal{L}=-\frac{1}{4}G^a_{\mu\nu}G^{a\mu\nu}+\sum^n_{i=1}[\bar{q_i} i\slashed{D}q_i-(m_i q^\dagger_{Li}q_{Ri}+h.c)]+\theta \frac{g^2_s}{32\pi^2}G^a_{\mu\nu}\tilde{G}^{a\mu\nu},
\end{equation}
where the dual of the gluon field strength tensor is,
\begin{equation}
\tilde{G}^{\mu\nu}=\frac{1}{2}\epsilon^{\mu\nu\gamma\delta}G_{\gamma\delta}.
\end{equation}
The last term in the QCD Lagrangian violates the discrete symmetries $P, T, CP$. Since all the quark masses are non-zero, the $\theta$ term in the Lagrangian must be present. The QCD depends on $\theta$ through some combination of parameters, $\bar{\theta}=\theta+\rm{arg(det}(M))$, where $M$ is the quark mass matrix \cite{adler,bell}. The neutron electric dipole moment (EDM) depends on $\bar{\theta}$ and from chiral perturbation theory we can obtain the neutron EDM as $d_n\simeq \rm{few}\times 10^{-16}\bar{\theta}\rm{e.cm}$. However the current experimental constraint on the neutron EDM is $d_n<\rm{few}\times 10^{-26}\rm{e.cm}$, which implies $\bar{\theta}\lesssim 10^{-10}$ \cite{baker}. The smallness of $\bar{\theta}$ is called the strong CP problem. To solve this, Peccei and Quinn, in 1977 \cite{quinn}, came up with an idea that $\bar{\theta}$ is not just a parameter but it is a dynamical field driven to zero by its own classical potential. They postulated a global $U_{PQ}(1)$ quasi symmetry which is a symmetry at the classical level but explicitly broken by the non perturbative QCD effects which produces the $\theta$ term, and spontaneously broken at a scale $f_a$. Thus the pseudo-Nambu-Goldstone bosons appear and these are known as the axions. The QCD axion mass $(m_a)$ is related to the axion decay constant $(f_a)$ by $m_a=5.7\times 10^{-12}\rm{eV}\Big(\frac{10^{18}\rm{GeV}}{f_a}\Big)$. So if we need axion decay constant less than the Planck scale $(M_{pl})$ then the mass of the axion is $m_a\gtrsim 10^{-12}\rm{eV}$ \cite{grilli}. Also there are other pseudo scalar particles which are not the actual QCD axions, but these particles have many similar properties like the QCD axions. These are called the axion like particles (ALPs). For ALPs, the mass and decay constant are independent of each other. These ALPs are motivated from the string theory \cite{sv}. 
The interaction of ALPs with the standard model particles is governed by the Lagrangian \cite{profumo}
\begin{equation}
\mathcal{L}=\frac{1}{2}\partial_\mu a\partial^\mu a-\frac{\alpha_s}{8\pi}g_{ag}\frac{a}{f_a}G^{\mu\nu}_a \tilde{G}^a_{\mu\nu}-\frac{\alpha}{8\pi}g_{a\gamma}\frac{a}{f_a}F^{\mu\nu}\tilde{F}_{\mu\nu}+\frac{1}{2}\frac{1}{f_a}g_{af}\partial_\mu a \bar{f}\gamma^\mu\gamma_5 f,
\label{eq:ak1}
\end{equation}	
where $g$'s are the coupling constants which depend on the model. The first term is the dynamical term of ALPs. The second, third, and last terms denote the coupling of ALPs with the gluons, photons, and fermion fields respectively. ALPs couple with the SM particles very weakly because the couplings are suppressed by $\frac{1}{f_a}$, where $f_a$ is called the axion decay constant and for ALPs, it generally takes larger value.

There is no direct evidence of axions in the universe. However, there are lots of experimental and astrophysical bounds on axion parameters. There are some ongoing searches for solar axions which correspond to $f_a\sim 10^7\rm{GeV}$ having sub-eV masses \cite{inoue,arik}. If solar axions were there, then it would violate the supernova 1987A result which requires $f_a\gtrsim 10^9\rm{GeV}$. Axions with $f_a\lesssim 10^8\rm{GeV}$ provide the component of hot dark matter\cite{mirizzi,mena,hann}. Large value of $f_a$ is allowed in the anthropic axion window and can be studied by isocurvature fluctuations \cite{hamann}. The laboratory bounds for the axions are discussed in \cite{y,cameron,robilliard,chou,si,kim,cheng,rosen}. The cosmological bounds for the cold axions produced by the vacuum realignment mechanism are discussed in\cite{hertz,visinelli}. The bounds on axion mass and decay constant are discussed in \cite{shellard,kawasaki,chang}, if cold axions are produced by the decay of axion strings.

Explaining the nature of dark matter and the dark energy is a major unsolved problem in modern cosmology. An interesting dark matter model is fuzzy dark matter (FDM)\cite{hu, hui}. The FDM are axion like particles (ALPs) with mass$(10^{-21}eV-10^{-22}eV)$ such that the associated de Broglie wavelength is comparable to the size of the dwarf galaxy $(\sim 2kpc)$. Axions and ALPs can be possible dark matter candidates \cite{duffy} or can be dynamical dark energy \cite{pradler}. Axions can also form clouds around black hole or neutron star from superradiance instabilities and change the mass and spin of the star \cite{bau,day}. Cold FDM can be produced by an initial vacuum misalignment and, to have the correct relic dark matter density, the axion decay constant should be $f_a \sim 10^{17} GeV$  \cite{hui}. This ultra light FDM was introduced to solve the cuspy halo problem. 

ALPs are pseudo-Nambu Golstone bosons which have a spin-dependent coupling with nucleons so that, in an unpolarized macroscopic body, there is no net long range field for ALPs outside the body. However, if the ALPs also have a CP violating coupling, then they can mediate long range forces even in unpolarized bodies \cite{Moody:1984ba, Raffelt:2012sp}. 

It has been pointed out recently \cite{hook} that if a compact star is immersed in an axionic potential (which will take place if the ALPs are FDM candidates), a long range field is developed outside the star.

The ALPs can be sourced by compact binary systems such as neutron star-neutron star (NS-NS), neutron star-white dwarf (NS-WD), and can have very small mass $(<10^{-19}eV)$. They can be possible candidates of FDM. 
The FDM density arises from a coherent oscillation of an axionic field in free space. If such axionic FDM particles have a coupling with nucleons, then the compact objects (NS, WD) immersed in the dark matter potential develop long range axionic hair. When such compact stars are in a binary orbit, they can lose orbital period by radiating the axion hair in addition to the gravitational wave\cite{hook,mohanty}.

In this paper, we study a model of ALPs sourced by the compact stars and put bounds on $f_a$ from the observations of the orbital period decay of compact binaries.

The paper is organised as follows. In section II, we compute in detail the axionic charge (including GR corrections) of compact stars immersed in a (ultra) low-mass axionic background potential. In section III, we show how the axionic scalar Larmor radiation can change the orbital period of compact binary systems. There may also be an axion mediated long ranged fifth force between the stars in a binary system. In section IV, we put constraints on $f_a$ for four compact binaries: PSR J0348+0432\cite{first}, PSR J0737-3039\cite{wer}, PSR J1738+0333\cite{third}, and PSR B1913+16 (Hulse Taylor Binary)\cite{hulse,mohanty}, available in the literature. In section V, we discuss the implication of the ALPs sourced by the compact binaries as the FDM. Finally, we summarize our results.\\
We use the units $\hslash=c=1$ throughout the paper.

\section{The axion profile for an isolated neutron star/white dwarf}
The axion Lagrangian at the leading order of $1/f_a$ is given in Eq.~(\ref{eq:ak1}). The axion pseudo shift symmetry $a\rightarrow a+\delta$ is used to remove the QCD theta angle. Suppose the fermions are quarks  and we give a chiral rotation to the quark field, so that only the non derivative coupling appears through the quark mass term. Such a field redefinition allow us to move the non derivative couplings into the two lightest quarks and all other quarks are integrated out. So we can work in the effective 2 flavour theory. Thus, in the chiral expansion, all the non derivative dependence of axion is contained in the pion mass term of the Lagrangian 
\begin{equation}
\mathcal{L}\supset 2B_0\frac{f^2_\pi}{4}<U M^\dagger_a+M_a U^\dagger>,
\end{equation}
where $U=e^\frac{{i\Pi}}{f_\pi}$ and $\Pi=\begin{bmatrix}
\pi^0& \sqrt{2}\pi^+\\
\sqrt{2}\pi^- & -\pi^0
\end{bmatrix}$, $B_0$ is related with the chiral condensate and it is determined by the pion mass term. $f_\pi$ is called the pion decay constant. We can obtain the effective axion potential from the neutral pion sector. On the vacuum, the neutral pion attains a vacuum expectation value and trivially be integrated out leaving the effective potential \cite{qcd}
\begin{equation}
V \approx  - m^2_\pi f^2_\pi\sqrt{1-\frac{4m_um_d}{(m_u+m_d)^2}\sin^2 \Big(\frac{a}{2f_a}\Big)},
\end{equation}
where $m_u$ and $m_d$ are the up and down quark masses respectively, and $m_\pi$ is the mass of pion.

It has been pointed out in \cite{hook} that, if we consider ALPs which couple to nucleons, then compact stars such as neutron stars and white dwarfs can be the source of long range axionic force. The reason for this long range force is as follows. In the vacuum, the potential for the ALPs is
\begin{equation}
V \approx  -\epsilon m^2_\pi f^2_\pi\sqrt{1-\frac{4m_um_d}{(m_u+m_d)^2}\sin^2 \Big(\frac{a}{2f_a}\Big)}.
\label{eq:kappa}
\end{equation} 
For simplicity we choose $m_u=m_d$ and, therefore, the mass of the ALPs in vacuum becomes
\begin{equation}
m_a=\frac{m_\pi f_\pi}{2f_a}\sqrt{\epsilon}.
\label{eq:kap}
\end{equation}
Inside a compact star, the quark masses are corrected by the nucleon density and the potential inside the star changes to
\begin{equation}
V=-m^2_\pi f^2_\pi \Big\{\Big(\epsilon-\frac{\sigma_N n_N}{m^2_\pi f^2_\pi}\Big) \Big|\cos\Big(\frac{a}{2f_a}\Big)\Big|+\mathcal{O} \Big(\Big(\frac{\sigma_N n_N}{m^2_\pi f^2_\pi}\Big)^2\Big)\Big\},
\label{eq:sig}
\end{equation}
and
\begin{equation}
\sigma_N=\sum_{q=u,d} m_q\frac{\partial m_N}{\partial m_q},
\end{equation}
where $n_N$ is the nucleon number density, $m_q$ is the quark mass, $m_\pi$ is the pion mass, and $f_\pi$ is the pion decay constant. $\sigma_N\sim 59MeV$ from lattice simulation \cite{Alarcon} and we consider the parameter space where $\epsilon\leq 0.1$\cite{hook}. The tachyonic mass of the ALPs is the square root of the second derivative of the potential Eq.~(\ref{eq:sig}) at $a=0$. Inside of the neutron star, $\sigma_N n_N/m^2_\pi f^2_\pi$ is not equal to zero  and $m_T\gtrsim m_a$.
Thus the magnitude of the tachyonic mass of the ALPs inside the compact star becomes
\begin{equation}
m_T=\frac{m_\pi f_\pi}{2f_a}\sqrt{\frac{\sigma_N n_N}{m^2_\pi f^2_\pi}-\epsilon},\hspace{1cm}r<r_{NS},
\label{eq:pa}
\end{equation}
where $r_{NS}$ is the radius of the compact star. The compact star can be the source of ALPs if its size is larger than the critical size given by \cite{hook}
\begin{equation}
r_c\gtrsim \frac{1}{m_T}.
\label{eq:tach}
\end{equation}  
For a typical neutron star and white dwarf, the condition Eq.~(\ref{eq:tach}) is satisfied. By matching the axionic field solution inside and outside of the compact star, we get the long range behaviour of the axionic field.
The axionic potential has degenerate vacua and this degeneracy can be weakly broken by higher dimensional operators suppressed by the Planck scale \cite{deg}. The degeneracy can also be broken by a finite density effect like the presence of a NS and WD. At the very high nuclear density, the axionic potential changes its sign which allows the ALPs to be sourced by the compact stars. Due to the very small size of the nuclei, it cannot be the source of the ALPs and long range axion fields arise only in large sized objects like NS and WD.

Using Eq.~(\ref{eq:kap}) in Eq.~(\ref{eq:pa}) we can write the tachyonic mass as 
\begin{equation}
m^2_T=\sigma_Nn_N/4f^2_a-m^2_a
\end{equation} 
Putting values of all the parameters and $m_a\sim 10^{-19}\rm{eV}$, we get the upper bound of the axion decay constant (using Eq.~(\ref{eq:tach})) as $f_a\lesssim 2.636\times 10^{17}\rm{GeV}$. Axions can never be sourced by neutron star if $f_a$ is greater than this upper bound. Similarly, white dwarf cannot be the source of axions if $f_a\gtrsim 9.95\times 10^{14}\rm{GeV}$.

Compact stars with large nucleon number density can significantly affect the axion potential. The second derivative of the potential Eq.~(\ref{eq:sig}) with respect to the field value is
\begin{equation}
\frac{\partial ^2 V}{\partial a^2}=m^2_\pi f^2_\pi \Big\{\Big(\epsilon-\frac{\sigma_N n_N}{m^2_\pi f^2_\pi}\Big)\frac{1}{4f^2_a}\cos\Big(\frac{a}{2f_a}\Big)+\mathcal{O} \Big(\Big(\frac{\sigma_N n_N}{m^2_\pi f^2_\pi}\Big)^2\Big)\Big\}.
\end{equation} 
Outside of the compact star, $\sigma_N=0$ which implies that 
\begin{equation}
\frac{\partial ^2 V}{\partial a^2}=m^2_\pi f^2_\pi \Big\{\epsilon\frac{1}{4f^2_a}\cos\Big(\frac{a}{2f_a}\Big)+\mathcal{O} \Big(\Big(\frac{\sigma_N n_N}{m^2_\pi f^2_\pi}\Big)^2\Big)\Big\}.
\end{equation}
Therefore, outside of the compact star $(r>r_{NS})$, the potential attains minima $(\frac{\partial ^2 V}{\partial a^2}>0)$ corresponding to the field values $a=0,\pm 4\pi f_a,...$ and maxima $(\frac{\partial ^2 V}{\partial a^2}<0)$ corresponding to the field values $a=\pm 2\pi f_a, \pm 6\pi f_a...$ etc.

Inside of the compact star $(r<r_{NS})$, $\sigma_N\neq 0$ 
and $\frac{\sigma_N n_N}{m^2_\pi f^2_\pi}>\epsilon$. Therefore, inside of the compact star, the potential has maxima at $a=0,\pm 4\pi f_a,...$ and minima at the field values $a=\pm 2\pi f_a, \pm 6\pi f_a...$ etc.

The axionic field becomes tachyonic inside of a compact star and reside on one of the local maxima of the axionic potential and, outside of the star, the axionic field rolls down to the nearest local minimum and stabilizes about it. The axionic field asymptotically reaches zero value $a=0$ at infinity. Therefore, throughout interior of the compact star the axionic field assumes a constant value $a=4\pi f_a$, the nearest local maximum. 

For an isolated compact star of constant density the equation of motion for the axionic field is \cite{hook} 
\begin{equation}
\nabla^{\mu}\nabla_{\mu} \left(\frac{\theta}{2}\right)= \begin{cases} 
     -m_T^2\sin \left(\frac{\theta}{2}\right) \text{sgn}\lbrace \cos \left(\frac{\theta}{2}\right)\rbrace & (r<r_{NS}),\\
       m_a^2\sin \left(\frac{\theta}{2}\right) \text{sgn}\lbrace \cos \left(\frac{\theta}{2}\right)\rbrace & (r>r_{NS}),
   \end{cases}
   \label{eq:axion_eom}
\end{equation} 
where $\theta=a/f_a$. The sgn function is required to take care of the absolute value $\vert \cos (\theta/2)\vert$ in the potential. Note that the equation of motion for the axionic field inside the compact star is satisfied by the field value $a=4\pi f_a$.

Assuming the exterior spacetime geometry due to the compact star to be the Schwarzschild, 
the axionic field equation Eq.~(\ref{eq:axion_eom}) becomes
\begin{equation}
\left(1-\frac{2GM}{r}\right)\frac{d^2a}{dr^2}+\frac{2}{r}\left(1-\frac{GM}{r}\right)\frac{da}{dr}=m_a^2a,
\label{eq:axion_eom_out}
\end{equation}       
where $M$ is the mass of the compact star, $G$ is the Newton's gravitational constant and we have used the approximation $\sin(\theta/2)\approx \theta/2$ for small $\theta$. 

At a large distance ($r>>2GM$) from the compact star, the axionic field Eq.~(\ref{eq:axion_eom_out}) becomes
\begin{equation}
\frac{d^2a}{dr^2}+\frac{2}{r}\frac{da}{dr}=m_a^2a
\label{eq:axion_eom_out_ld}.
\end{equation}
Assuming $a=\xi(r)/r$, the above equation reduces to $\xi''-m_a^2\xi=0$ (where prime denotes derivative with respect to $r$). This has the solution $\xi= C_1e^{m_ar}+C_2e^{-m_ar}$. Since $a\rightarrow 0$ in the limit $r\rightarrow \infty$, $C_1=0$. Thus, $a$ behaves as $a\sim q_{eff} e^{-m_a r}/r$ where we rename the integration constant $C_2$ as $q_{eff}$. Further, for sufficiently light mass ($m_a<<1/D<<1/r_{NS}$ where $D$ is the distance between the stars in a binary system), the scalar field has a long range behaviour with an effective charge $q_{eff}$. For scalar Larmor radiation, the orbital frequency ($\omega$) of the binary pulsar should be greater than the mass of the particle that is radiated (i.e. $\omega>m_a$). This translates the mass spectrum of radiated ALPs for a typical neutron star- neutron star (NS-NS) or a neutron star- white dwarf (NS-WD) binary system into $m_a\lesssim 10^{-19}$ eV. Also, the axion Compton wavelength should be much larger than the binary distance in order to use the massless limit in the computation of scalar radiation and effective charge, i.e. $m_a^{-1}>> D$. The critical value of axion mass required for the scalar radiation and the binary distance for four compact binary systems are given in Table \ref{tableI} which is consistent with the assumption of $m_a\lesssim 10^{-19}eV$. Consequently, the axion Compton wavelength (inverse of axion mass) is larger than the binary distance and, hence, the size of the star (size of NS is $10^{20}\rm{GeV^{-1}}$ and size of WD is $10^{23}\rm{GeV^{-1}}$).

\begin{table}
\caption{\label{tableI} Summary of the axion Compton wavelength $(m^{-1}_a)$ and binary distance $D$ for all the four compact binaries. All relevant parameters for the numerical calculation are given in section~\ref{secIV}. }
\begin{tabular}{ lcc  }
 \hline
Binary system & Critical value of $m^{-1}_a$ $(GeV^{-1})$& binary separation D$(GeV^{-1})$  \\
 \hline
PSR J0348+0432  &  $ 2.14\times 10^{27}$ & $4.64\times 10^{24}$\\
 PSR J0737-3039 & $ 2.08\times 10^{27}$ & $4.83\times 10^{24}$\\
 PSR J1738+0333 &  $ 7.41\times 10^{27}$& $9.65\times 10^{24}$\\
PSR B1913+16 &  $ 6.76\times 10^{27}$& $1.08
\times 10^{25}$\\
 \hline
\end{tabular} 
\end{table}

\begin{figure}[!htbp]
\centering
\subfigure[$V$ vs. $a$]{\includegraphics[width=3.5in,angle=360]{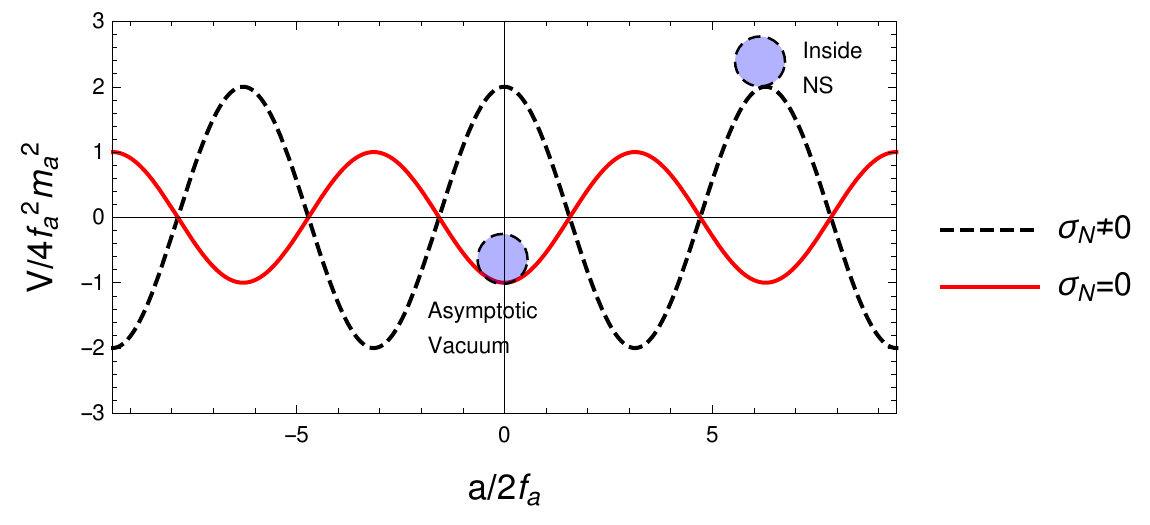}\label{subfig:vvsa}}
\subfigure[$V$ vs. $r$]{\includegraphics[width=3.5in,angle=360]{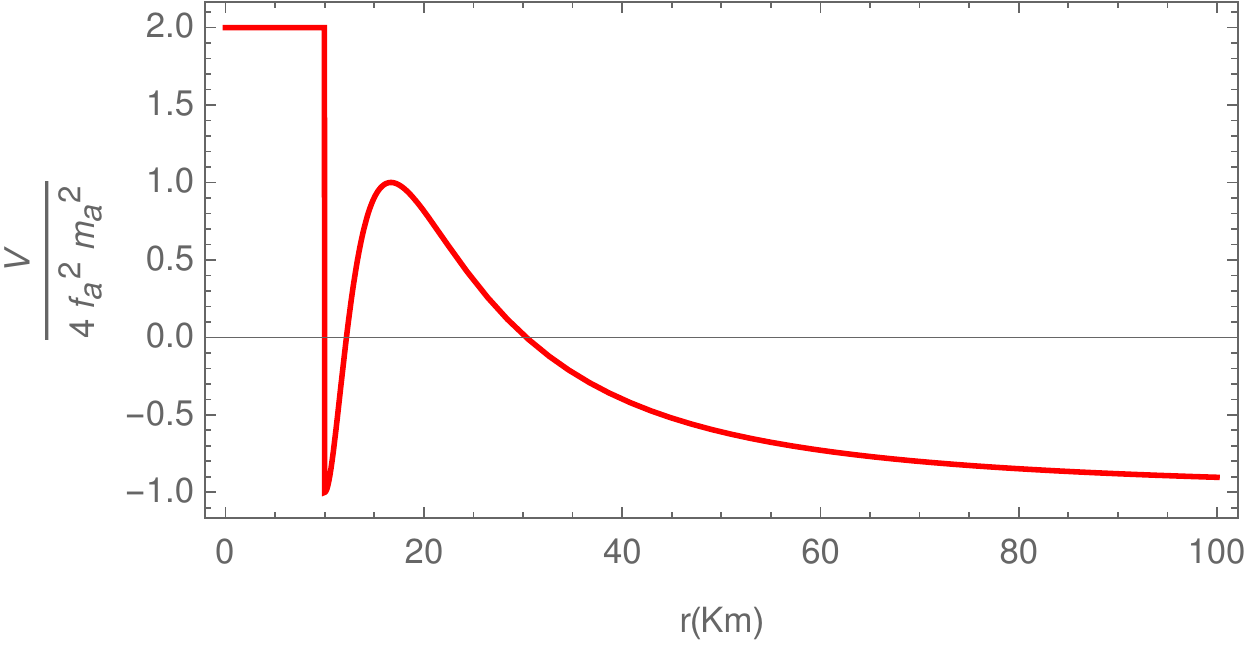}\label{subfig:vvsr}}
\subfigure[$a$ vs. $r$]{\includegraphics[width=3.5in,angle=360]{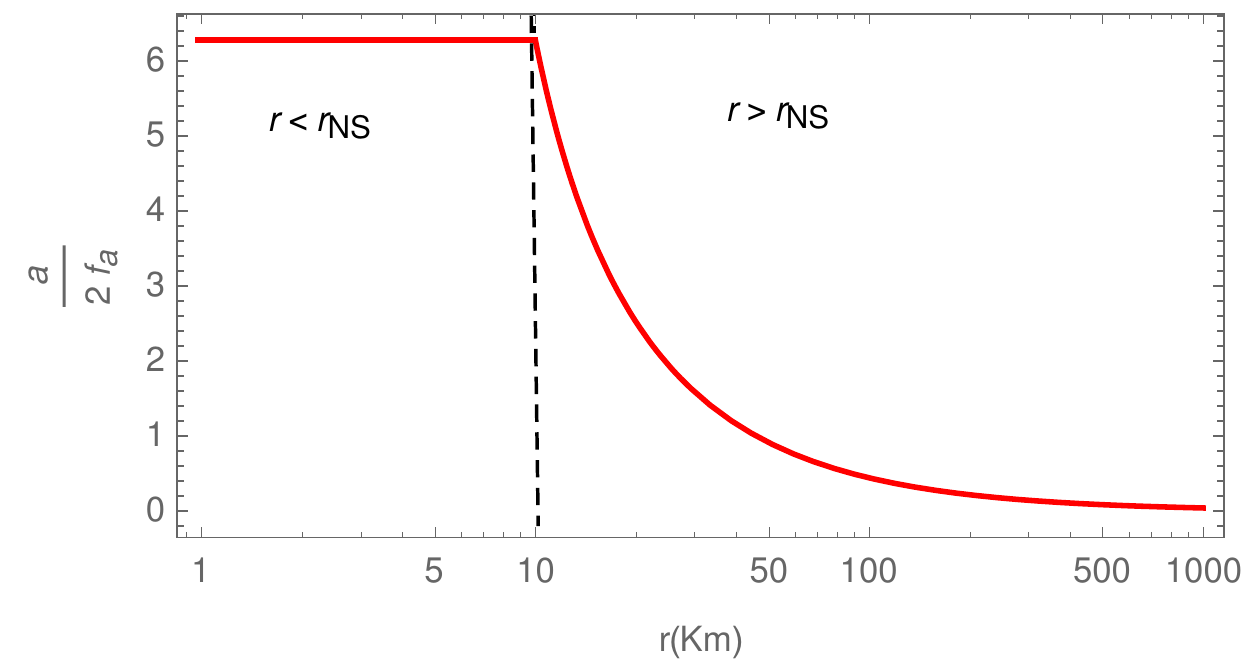}\label{subfig:avsr}}
\caption{(a)plot of the axionic potential V as the function of the axionic field. We assume $m_T^2/m_a^2=2$. The black dashed line corresponds to $\sigma_N\neq 0$ (i.e; inside the compact star) and the red solid line corresponds to  $\sigma_N= 0$. Note that the axionic field evolves from the local maximum $a=4\pi f_a$ inside a compact star to nearest local minimum $a=0$ outside the compact star. (b)The plot of $V$ as the function of $r$ inside and outside of the neutron star. Note that there is discontinuity in $V(r)$ at $r=r_{NS}$ due to sign change in the potential. (c)plot of the axionic field $a$ as the function of $r$. We assume neutron star as the example of the compact object in the plots. The typical mass and radius of a neutron star are $M=1.4M_{\odot}$ and $r_{NS}=10km$ respectively. The similar type of profiles we can obtain for white dwarfs.}
\label{fig:axion_profile}
\end{figure}

\begin{figure}[h]
\includegraphics[height=8cm]{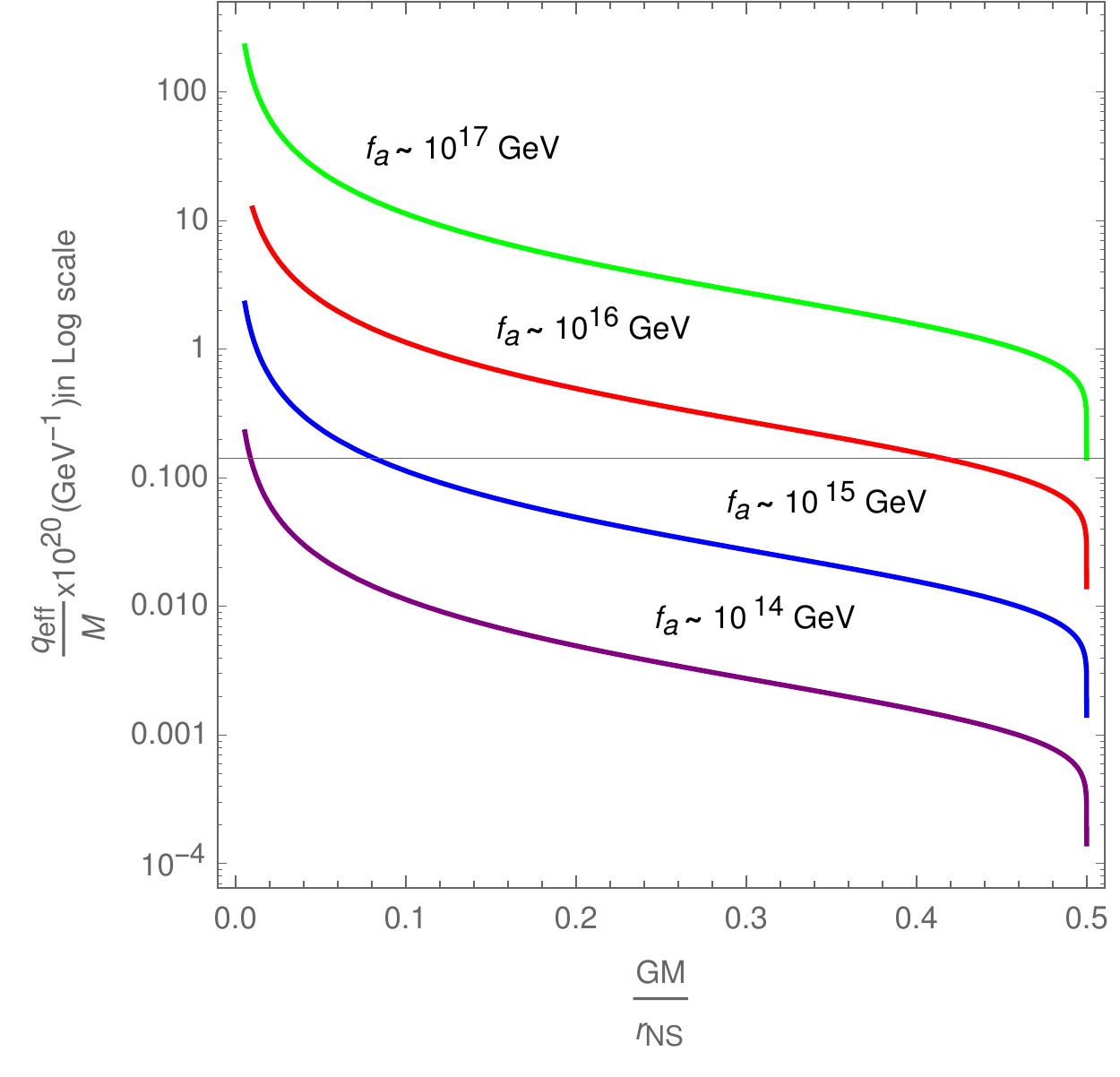}
\caption{ The variation of effective charge to mass ratio of the neutron star with the ratio of mass to radius for different values of axion decay constant.  }
\label{fig:axion_y}
\end{figure}

To identify the effective charge $q_{eff}$, we exploit the continuity of the axion field across the surface of the compact star. Therefore, we solve Eq.~(\ref{eq:axion_eom_out}) in the massless limit ($m_a\rightarrow 0$), i.e. 
\begin{equation}
\left(1-\frac{2GM}{r}\right)\frac{d^2a}{dr^2}+\frac{2}{r}\left(1-\frac{GM}{r}\right)\frac{da}{dr}=0.
\label{eq:axion_eom_out_massless}
\end{equation}
Integrating Eq.~(\ref{eq:axion_eom_out_massless}) we get $a'=-C_3/{r^2(1-2GM/r)}$ and further integration yields $a=-\frac{C_3}{2GM}\ln\left(1-2GM/r\right)+C_4$, where $C_3$ and $C_4$ are integration constants. For $r>>2GM$ limit, $a\rightarrow q_{eff}/r$ and, therefore, $C_3=q_{eff}$ and $C_4=0$. Therefore, we get the axionic field profile outside the compact star
\begin{equation}
a=-\frac{q_{eff}}{2GM}\ln\left(1-\frac{2GM}{r}\right).
\label{eq:axion_out_massless}
\end{equation}
The behaviour of the axionic potential as a function of the axionic field and distance are illustrated in FIG.\ref{fig:axion_profile}. The nature of the axionic field as we go from inside to outside of a compact star is also shown in FIG.\ref{fig:axion_profile}. Variation of the effective charge to mass ratio of a compact star is shown in FIG.\ref{fig:axion_y} as a function of the mass to radius ratio for different decay constants.

\begin{figure}
\centering
\includegraphics[height=6cm]{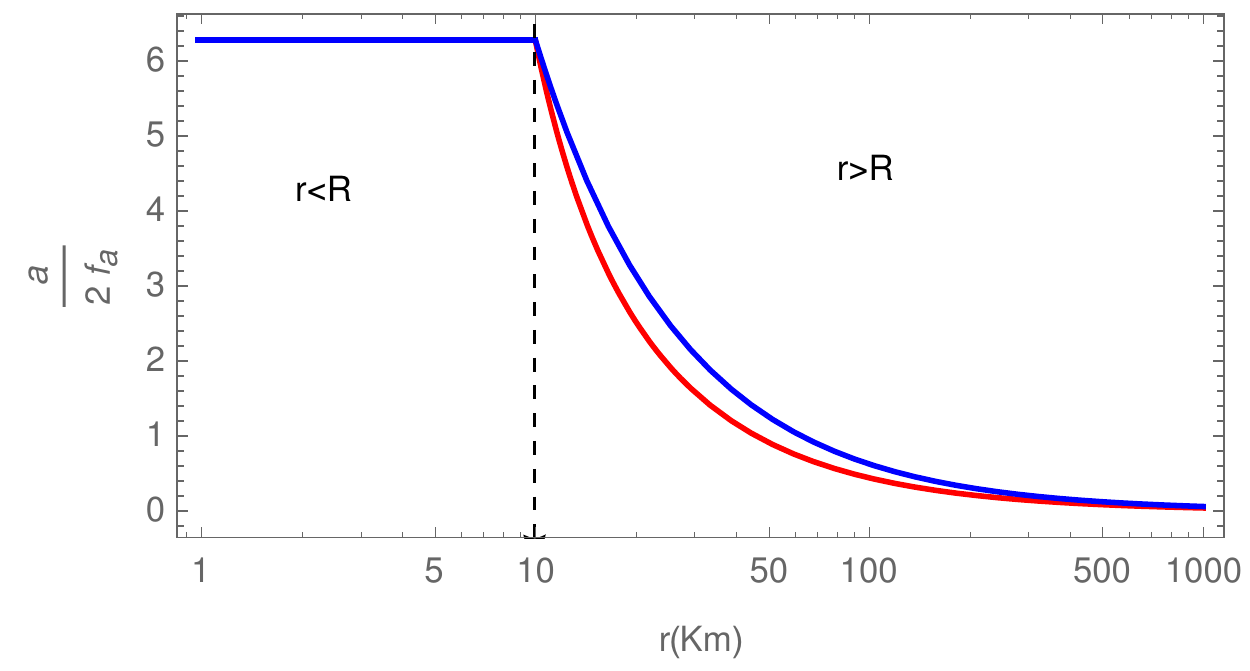}
\caption{plot of the axion field $a$ as the function of $r$. The blue curve stands for the axion field $a\sim q_{eff}/r$ and the red curve stands for the axion field $a\sim -q_{eff}/2GM \ln(1-2GM/r)$ outside of the neutron star. For blue curve, the effective axion charge is $q_{eff}=4\pi f_a r_{NS}$ and for the red curve $q_{eff}$ is given by Eq.\ref{eq:axion_e}.}
\label{fig:aaxion}
\end{figure}

At the surface of the compact star, $a(r_{NS})=4\pi f_a$. Thus we identify
\begin{equation}
q_{eff}=-\frac{8\pi GM f_a}{\ln\left(1-\frac{2GM}{r_{NS}}\right)}.
\label{eq:axion_e}
\end{equation}
If $\frac{GM}{r_{NS}}<<1$, $q_{eff}\sim 4\pi f_a r_{NS}$ \cite{hook}. However, for a typical neutron star ($M=1.4M_{\odot}$ and $r_{NS}= 10$ km) the above correction is not negligible. For white dwarf the effect is negligible. The charges can be both positive as well as negative depending on the sign of the  axionic field values at the surface of the compact star. If $q_1$ and $q_2$ are the charges of two compact stars, then if $q_1q_2>0$ the two stars attract each other and, if $q_1q_2<0$, then they repel each other \cite{hook}. For neutron star, the new effective axion charge Eq.~(\ref{eq:axion_e}) is smaller than $4\pi f_a r_{NS}$ by $21.46\%$. The effect of new axion charge is illustrated in Fig.~\ref{fig:aaxion} where the plot of axion profile inside and outside of a neutron star is shown. 

\section{Axionic fifth force and the scalar radiation for the compact binaries}
Such a long range axionic field mediates a ``fifth" force (in addition to the Newtonian gravitational force) between the stars of a binary system (NS-NS or NS-WD),
\begin{equation}
F_5=\frac{q_1 q_2}{4\pi D^2},
\end{equation}  
where $q_{1,2}$ are effective charges of the stars in the binary system. Due to the presence of this scalar mediated fifth force the Kepler's law is modified by\cite{jana}
\begin{equation}
\omega^2=\frac{G(m_1+m_2)}{D^3}(1+\alpha),
\label{eq:kepler}
\end{equation}
where $\alpha=\frac{q_1q_2}{4\pi Gm_1m_2}$ is the ratio of the scalar mediated fifth force to the gravitational force, $\omega$ is the angular frequency of orbital motion of the stars, $m_1$ and $m_2$ are the masses of the stars and $\mu=m_1m_2/(m_1+m_2)$ is the reduced mass of the binary system. There are constraints on the fifth force from either scalar-tensor theories of gravity \cite{liu,jana,alexander} or the dark matter components \cite{croon,laha,alexander}. In this paper we show that the constraint on $\alpha$ from time period loss by scalar radiation is more stringent than the measured change in orbital period Eq.~(\ref{eq:kepler}) due to fifth force.

The orbital period of the binary star system decays with time because of the energy loss primarily due to the gravitational quadrupole radiation and about one percent due to ultra light scalar or pseudoscalar Larmor radiation. The total power radiated for such quasi-periodic motion of a binary system is 
\begin{equation}
\frac{dE}{dt}=-\frac{32}{5}G\mu^2D^4\omega^6(1-e^2)^{-\frac{7}{2}}\Big(1+\frac{73}{24}e^2+\frac{37}{96}e^4\Big)-\frac{\omega^4 p^2}{24\pi}\frac{(1+e^2/2)}{(1-e^2 )^{\frac{5}{2}}},
\label{eq:axion_t}
\end{equation}
where $e$ is the eccentricity of the elliptic orbit and $E$ is the total energy of the binary system.
The first term on the r.h.s. is the gravitational quadrupole radiation formula \cite{mohanty,croon} and the second term is the massless scalar dipole radiation formula \cite{mohanty,hook,krause}. There is the radiation of the ALPs if the orbital frequency is greater than the mass of the ALPs. The dipole moment in the centre of mass frame of the binary system can be written as
\begin{equation}
p=q_1r_1-q_2r_2=q_1 \frac{\mu D}{m_1}-q_2\frac{\mu D}{m_2},
\end{equation}
or,
\begin{equation}
p=8\pi G f_a\mu D\left[\frac{1}{\ln\left(1-\frac{2Gm_2}{r_{NS}}\right)}-\frac{1}{\ln\left(1-\frac{2Gm_1}{r_{NS}}\right)}\right],
\label{eq:master}
\end{equation}
where $r_{1,2}$ are the radial distances of the stars in the binary system from the centre of mass along the semi-major axis. For nonzero scalar radiation the charge-to-mass ratio ($q/m$) should be different for two stars. Thus for the companion star in a binary system with the equal effective charge, there should be some mass difference of the two stars. The decay of the orbital time period is given by\cite{peters,mohanty}
\begin{equation}
\dot{P_b}=6\pi G^{-\frac{3}{2}}(1+\alpha)^{-\frac{3}{2}}(m_1m_2)^{-1}(m_1+m_2)^{-\frac{1}{2}}D^{\frac{5}{2}}\Big(\frac{dE}{dt}\Big),
\label{eq:axion_n}
\end{equation}
where $P_b=2\pi/\omega$. NS-NS binaries (with different mass components) as well as NS-WD binaries are the sources for the scalar Larmor radiation and also for the axion mediated fifth force. On the other hand, NS-BH systems can be the source of scalar radiation but there is no long range fifth force in between, as the scalar charges for the black holes (BH) are zero\cite{mairi}.

In the next section, we consider four compact binaries and put constraints on $f_a$.

\section{Constraints on axion parameters of different compact binaries}
\label{secIV}
\subsection{PSR J0348+0432}
This binary system is consist of a neutron star and a low mass white dwarf companion. The orbital period of the quasi-periodic binary motion is $P_b=2.46h$. The mass of the neutron star in this binary system is $M_p=2.01 M_\odot$ and the mass of the white dwarf is $M_{WD}=0.172M_\odot$. The radius of the white dwarf is $r_{WD}=0.065 R_{\odot}$ and we assume the radius of the neutron star $r_{NS}=10 km$. We compute the semi-major axis of the orbit using Kepler's law Eq.~(\ref{eq:kepler}). The observed decay of the orbital period is $\dot{P_b}=0.273\times 10^{-12}ss^{-1}$\cite{first}. This is primarily due to gravitational quadrupole radiation from the binary NS-WD system. The contribution from the radiation of some scalar or pseudoscalar particles must be within the excess of the decay of the orbital period, i.e. $\dot{P}_{b(scalar)}\leqslant \vert \dot{P}_{b(observed)}- \dot{P}_{b(gw)}\vert$. If ALPs are emitted as scalar Larmor radiation, then we can find the upper bound on the axion decay constant. Using Eqs.~(\ref{eq:kepler}),(\ref{eq:axion_t}), (\ref{eq:master}) and (\ref{eq:axion_n}) and taking the ALPs as massless, we obtain an upper bound on the axion decay constant as, $f_a\lesssim 1.66\times 10^{11}$ GeV. The ratio of the axionic fifth force and the Newtonian gravitational force between the stars in this system comes out to be $\alpha\lesssim 5.73\times 10^{-10}$ .

\subsection{PSR J0737-3039}
It is a double neutron star binary system whose average orbital period is $P_b=2.4h$. Its observed orbital period decays at a rate $\dot{P_b}=1.252\times 10^{-12}ss^{-1}$.  The pulsars have masses $M_1=1.338M_\odot$ and $M_2=1.250M_\odot$. The eccentricity of the orbit is $e=0.088$\cite{wer}. Using Eqs.~(\ref{eq:kepler}), (\ref{eq:axion_t}), (\ref{eq:master}),  and (\ref{eq:axion_n})  we obtain the upper bound on the axion decay constant as $f_a\lesssim 9.76\times 10^{16}$ GeV. Beside the axion radiation, axion mediated fifth force arises in this binary system. We obtain the value of $\alpha\lesssim 9.21\times 10^{-3}$ .

\subsection{PSR J1738+0333}
This pulsar-white dwarf binary system has an average orbital period $P_b=8.5h$ and the orbit has a very low eccentricity $e<3.4\times 10^{-7}$. The mass of the pulsar is $M_p=1.46M_\odot$ and the mass of the white dwarf is $M_{WD}=0.181M_\odot$. The radius of the white dwarf is $r_{WD}=0.037R_{\odot}$. The rate of the intrinsic orbital period decay is $\dot{P_b}=25.9\times 10^{-15}ss^{-1}$\cite{third}. Using this system, we obtain the upper bound on the axion decay constant as $f_a\lesssim 2.03\times 10^{11}$ GeV. The value of $\alpha$ comes out $\lesssim 8.59\times 10^{-10}$.

\subsection{PSR B1913+16: Hulse Taylor binary pulsar}
The observed orbital period of the Hulse Taylor binary decays at the rate of $\dot{P_b}=2.40\times 10^{-12}ss^{-1}$. The masses of the stars in this binary system are $m_1=1.42  M_\odot$ and $m_2=1.4  M_\odot$\cite{mohanty} . The eccentricity of the orbit is $e=0.617127$ and the average orbital frequency is $\omega=0.2251\times 10^{-3}s^{-1}$. For this system, we obtain the upper bound on the decay constant as $f_a\lesssim 2.12\times 10^{17}$ GeV. We obtain the value of $\alpha$ for this system $\lesssim 3.4\times 10^{-2}$. Note that the binary orbit of this system is highly eccentric. As a result the contributions of the eccentricity factors in the radiation formulae Eq.~(\ref{eq:axion_t}) are important. For the GW radiation the eccentricity factor is 11.85 and, for the scalar radiation, it is 3.94.   

In Table \ref{tableII}, we have obtained the upper bound of the axion decay constant and the relative strength of axion mediated force for the four compact binaries.
\begin{table}[h]
\caption{\label{tableII} Summary of the upper bounds on the axion decay constant $f_a$ of ALPs radiated from compact binaries. For all the binaries we assume $m_a<10^{-19}$ eV.}
\centering
\begin{tabular}{ lcc  }
 
 \hline
Compact binary system & $f_a$ (GeV) & $\alpha$\\
 \hline
PSR J0348+0432  & $\lesssim 1.66\times 10^{11}$  & $\lesssim 5.73\times 10^{-10}$ \\
 PSR J0737-3039 & $\lesssim   9.76\times 10^{16}$  &$\lesssim 9.21\times 10^{-3}$ \\
 PSR J1738+0333 & $\lesssim 2.03\times 10^{11}$  & $\lesssim 8.59\times 10^{-10}$\\
PSR B1913+16 & $\lesssim 2.12\times 10^{17}$  & $\lesssim 3.4\times 10^{-2}$ \\
 \hline
\end{tabular}
\end{table}

\begin{thebibliography}{100} 
\bibitem{quinn} R.D. Peccei and H.R. Quinn, Phys. Rev. Lett. {\bf 38}, 1440-1443 (1977).
\bibitem{weinberg} S.Weinberg.,Phys.Rev.Lett 40 (1978)223-226
\bibitem{wil} F.Wilczek, Phys.Rev.Lett 40 (1978) 279-282.
\bibitem{peccei}R. D. Peccei and H. R. Quinn, Phys.Rev D {\bf 16}, 1791 (1977).
\bibitem{adler} S.L.Adler., Phys.Rev 177 (1969) 2426-2438.
\bibitem{bell} J.S.Bell and R.Jackiw., Nuovo Cim A60(1969) 47-61
\bibitem{baker} C. A. Baker, D. D. Doyle, P. Geltenbort, K. Green, M. G. D. van der Grinten, P. G. Harris, P. Iaydjiev, S. N. Ivanov, D. J. R. May, J. M. Pendlebury, J. D. Richardson, D. Shiers, and K. F. Smith, Phys. Rev. Lett. {\bf 97}, 131801 (2006).
\bibitem{grilli} G. Grilli di Cortona, E. Hardy, J. Pardo Vega, and G. Villadoro, The QCD axion, precisely, JHEP 01 (2016) 034.
\bibitem{sv}P. Svrcek, E. Witten, JHEP 0606:051, 2006. 
\bibitem{profumo}S. Profumo, An introduction to particle dark matter, World Scientific.
\bibitem{inoue} Y.Inoue et al.,Phys.Lett.B 668(2008).
\bibitem{arik} E.Arik et al. JCAP 0902 (2009).JCAP 0507 (2005).
\bibitem{mirizzi} S.Hannestad, A.Mirizzi and G. Raffelt., JCAP 0507 (2005).
\bibitem{mena} A. Melchiorri, O. Mena, and A. Slosar, Phys. Rev. D {\bf 76}, 041303(R) (2007).
\bibitem{hann} S.Hannestad, A.Mirizzi, G.G. Raffelt and Y.Y.Y.Wong, JCAP 0804 (2008).
\bibitem{hamann} J.Hamann, S.Hannestad, G.G Raffelt and Y.Y.Y.Wong.,Isocurvature forecast in the anthropic axion window JCAP 0906 (2009) 022.
\bibitem{y} Y.Semertzidis et al 1990., Limits on the production of light scalar and pseudoscalar particles Phys.Rev.Lett.64 2988-91.
\bibitem{cameron} R.Cameron et al 1993., Search for nearly massless, weakly coupled particles by optical techniques Phys.Rev.D 47 3707-25.
\bibitem{robilliard} C. Robilliard, R. Battesti, M. Fouche, J. Mauchain, A.-M. Sautivet, F. Amiranoff, and C. Rizzo, Phys. Rev. Lett. {\bf 99}, 190403 (2007).

\bibitem{chou} A. S. Chou, W. Wester, A. Baumbaugh, H. R. Gustafson, Y. Irizarry-Valle, P. O. Mazur, J. H. Steffen, R. Tomlin, X. Yang, and J. Yoo, Phys. Rev. Lett. {\bf 100}, 080402 (2008).
\bibitem{si} P. Sikivie, D. B. Tanner, and K. van Bibber, Phys. Rev. Lett. {\bf 98}, 172002 (2007).

\bibitem{kim} J.E Kim, Phys. Rep. {\bf 150}, 1-177 (1987).

\bibitem{cheng}H-Y Cheng 1988., The strong CP problem revisited Phys.Rep.158.
\bibitem{rosen} L.J Rosenberg and K A van Bibber 2000., Phys.Rep.325.
\bibitem{hertz} M. P. Hertzberg, M. Tegmark, and F. Wilczek, Phys. Rev. D 78, 083507 (2008).

\bibitem{visinelli} L. Visinelli and P. Gondolo, Phys. Rev. D {\bf 80}, 035024 (2009).

\bibitem{shellard} R. A. Battye and E. P. S. Shellard, Phys. Rev. Lett. {\bf 73}, 2954 (1994); Erratum: Phys. Rev. Lett. 76, 2203 (1996).

\bibitem{kawasaki} M. Yamaguchi, M. Kawasaki, and J. Yokoyama, Phys. Rev. Lett. {\bf 82}, 4578 (1999).

\bibitem{chang} C. Hagmann, S. Chang, and P. Sikivie, Phys. Rev. D {\bf 63}, 125018 (2001).

\bibitem{hu} W. Hu, R. Barkana, and A. Gruzinov, Phys. Rev. Lett. {\bf 85}, 1158 (2000).
\bibitem{hui}L. Hui, J. P. Ostriker, S. Tremaine, and E. Witten, Phys. Rev. D {\bf 95}, 043541 (2017).
\bibitem{duffy}D. Duffy and K. van Bibber, New J. Phys. {\bf 11}, 105008 (2009).
\bibitem{pradler} M.Kamionkowski, J. Pradler, and D.G.E. Walker, Phys. Rev. Lett {\bf 113},251302 (2014).
\bibitem{bau}D. Baumann, H. S. Chia, and R. A. Porto, Phys. Rev. D 99, 044001 (2019).
\bibitem{day}F. V. Day, and J. I. Mcdonald, JCAP10(2019)051.
\bibitem{Moody:1984ba} 
  J.~E.~Moody and F.~Wilczek,
  Phys.\ Rev.\ D {\bf 30}, 130 (1984).
  doi:10.1103/PhysRevD.30.130.
\bibitem{Raffelt:2012sp} 
  G.~Raffelt,
  Phys.\ Rev.\ D {\bf 86}, 015001 (2012)
  doi:10.1103/PhysRevD.86.015001
  [arXiv:1205.1776 [hep-ph]].

\bibitem{hook} A. Hook and J. Huang, Journal of High Energy Physics (2018) 2018:36.
\bibitem{mohanty} S. Mohanty and P. K. Panda, Phys. Rev. D {\bf 53}, 5723 (1996).
\bibitem{first} J. Antoniadis et al., Science {\bf 340}, 6131 (2013).
\bibitem{wer}M. Kramer et al., Science {\bf 314}, 97 (2006).
\bibitem{third} P. C. C Freire et al., Mon. Not. R. Astron. Soc. {\bf 423}, 3328 (2012).
\bibitem{hulse}R.A. Hulse and J.H. Taylor, Ap. J. Lett 195, L51 (1975); J.H. Taylor and J.M. Weisberg, Ap. J. 253, 908 (1982); J.M. Weisberg and J.H. Taylor, Phys. Rev. Lett. 52, 1348 (1984).
\bibitem{qcd}G. G. D. Cortona, E. Hardy, J. P. Vega, and G. Villadoro, 10.1007/JHEP01(2016)034.
\bibitem{Alarcon} J. M. Alarcon, J. M. Camalich, and J.A. Oller, Phys. Rev. D {\bf 85},051503(R) (2012).
\bibitem{deg} M. Kamionkowsky and J. March-Russell, Phys. Lett.B282 (1992)137-141.
\bibitem{jana} S. Jana and S. Mohanty, Phys. Rev. D {\bf 99}, 044056 (2019).
\bibitem{liu} T. Liu, X. Zhang, and W. Zhao, Phys. Lett. B {\bf 777}, 286-293 (2018).
\bibitem{alexander} S. Alexander, E. McDonough, R. Sims, and N. Yunes, Class. Quant. Grav. {\bf 35}, 235012 (2018).
\bibitem{croon} D. Croon, A. E. Nelson, C. Sun, D. G. E. Walker, and Z.-Z. Xianyu, ApJ Lett. {\bf 858}:L2 (5pp), 2018.
\bibitem{laha} J. Kopp, R. Laha, T. Opferkuch, and W. Shepherd, Journal of High Energy Physics 1811 (2018) 096.
\bibitem{krause}D. E. Krause, H. T. Kloor, and E. Fischbach, Phys. Rev. D {\bf 49}, 6892 (1994).
\bibitem{peters}P. C. Peters and J. Mathew, Phys. Rev. {\bf 131}, 435 (1963).
\bibitem{mairi} J. Huang, M. C. Johnson, L. Sagunski, M. Sakellariadou, and J. Zhang, Phys. Rev. D {\bf 99}, 063013 (2019).
\bibitem{kari} K. Enqvist, R. J. Hardwick, T. Tenkanen, V. Vennin, and D. Wands, Journal of Cosmology and Astroparticle Physics 1802 (2018) 006.
\bibitem{er} T. Kobayashi, R. Murgia, A. De Simone, V. Irsic, and M. Viel, Phys. Rev. D
{\bf 96}, 123514 (2017).








\end{thebibliography}

\section{Implication for the axionic Fuzzy dark matter (FDM)}
The ALPs that are radiated from the compact binaries can be possible candidates of FDM whose mass is $\sim \mathcal{O}(10^{-21}eV-10^{-22}eV)$. 
At the very early universe the axionic field evolves with a cosine potential
\begin{equation}
V\Big(\frac{a}{f_a}\Big)= m^2_a f^2_a\Big[1-\cos\Big(\frac{a}{f_a}\Big)\Big].
\end{equation} 
The equation of motion for the axionic field is
\begin{equation}
\ddot{a}+3H\dot{a}-\frac{1}{R^2}\nabla^2 a+m^2_a a=0,
\label{fr}
\end{equation}
where $R(t)$ is the scale factor in the FRW spacetime. Taking the Fourier transform of Eq.~(\ref{fr}), the modes decouple and we have,
\begin{equation}
\ddot{a_k}+3H\dot{a_k}+\frac{k^2}{R^2} a_k+m^2_a a_k=0
\end{equation}
For non relativistic (small k) or zero modes, the third term becomes zero and the equation of motion of the axionic field is damped harmonic oscillatory. The axionic field takes a constant value as long as $H\gtrsim m_a$ which fixes the initial misalignment angle and then the axionic field starts oscillating with a frequency $\sim m_a$. When the oscillation starts at $H\sim m_a$, then the energy density of axionic field is of the order of $m^2_a a^2_0$, where $a_0$ is the initial field value during inflation. The oscillation modes are damped as $R^{-\frac{3}{2}}$. The energy density of the axionic field when it is oscillating, goes as $\frac{1}{R^3}$. Hence, at the late time, the axionic energy density redshifts like a cold dark matter. The ratio of dark matter to radiation energy densities increases as $\frac{1}{T}$ with the expansion of the universe and the dark matter starts dominating over radiation at $T\sim 1eV$. Using these facts the dark matter relic density becomes \cite{hui}
\begin{equation}
\Omega_{DM}\sim 0.1 \Big(\frac{a_0}{10^{17}GeV}\Big)^2\Big(\frac{m_a}{10^{-22}eV}\Big)^\frac{1}{2},
\end{equation}
where $a_0=\theta_0 f_a$ and $\theta_0$ is the initial misalignment angle which can take values in the range $-\pi<\theta_0<\pi$. Since the coupling of ALPs with matter is proportional to $\frac{1}{f_a}$, large values of $f_a$ corresponds to weaker coupling with matter. Therefore, direct detection of the ALPs in this scale is much more difficult. However, the ALPs in this large $f_a$ scale has some theoretical motivations\cite{sv}. Axion decay constant in the GUT scale implies that a single axion condensate can trigger the breaking of symmetries in nature. ALPs of mass $\mathcal{O}(10^{-21}eV-10^{-22}eV)$ sourced by the binary systems can give rise to the correct relic density of FDM if the axion decay constant is $f_a\sim 10^{17}GeV$ and initial misalignment angle $\theta_0\sim \mathcal{O}(1)$. Any value of $f_a$ other than $10^{17}GeV$ requires fine tuning of the initial misalignment angle which can take value from $-\pi$ to $+\pi$.
 

For the NS-WD binaries PSR J0348+0432 and PSR J1738+0333, the bound on the axion decay constant $(f_a\lesssim \mathcal{O}(10^{11} GeV))$ is well below the GUT scale and this gives the stronger bound. This implies that if the ultra-light ALPs has to be FDM then they do not couple with gluons.

\section{Conclusions and Discussions}
In this paper, we have obtained upper bounds on the decay constant of the ultralight ALPs from the study of decay in orbital period of the compact binary stars (NS-NS, NS-WD). Compact stars such as neutron stars and white dwarfs can be the source of ALPs. We have assumed that the mass of the ALPs is sufficiently low such that the axionic field has a long range behaviour over a distance between the binary companions. Due to such axionic field, the binary system will emit scalar Larmor radiation. Although the gravitational quadrupole radiation mainly contributes to the decay of orbital period, the contribution of scalar radiation is not negligible. However, its contribution must be within the excess value of the observed decay in the orbital period. For the NS-NS and NS-WD binary systems, an additional axionic ``fifth" force arises which is not relevant as much as the scalar radiation in our study. 

We have obtained the axionic profile for an isolated compact star assuming it to be a spherical object of uniform mass density. We have identified the form of effective axionic charge of the compact star\cite{hook} and its GR correction. We have also considered eccentricity of the orbit of binary system-- a generalization of previous result for axionic scalar radiation \cite{hook}. Using the updated formula for the total power radiated, we have studied four compact binary systems: PSR J0348+0432, PSR J0737-3039, PSR J1738+0333, and PSR B1913+16 (Hulse-Taylor binary pulsar). The upper bound on the axion decay constant $f_a$ is found as $f_a\lesssim \mathcal{O}(10^{11} GeV)$. 

If the mass of the ALPs which are sourced by compact binaries is $\mathcal{O}(10^{-21}eV-10^{-22}eV)$ and $f_a\sim 10^{17} GeV$, then they can contribute to relic density of FDM. However, the bound $f_a \lesssim \mathcal{O}(10^{11} GeV)$ from WD binaries do not favour ALPs as the FDM.

 ALPs  can give rise to isocurvature fluctuations during inflation which are tightly constrained from CMB observation. The Hubble scale during inflation (for single field slow roll models) is $H_I = 8\times 10^{13} \sqrt{
r/0.1} GeV $\cite{kari} and, therefore, for our bound  $f_a\lesssim \mathcal{O}(10^{11} GeV)$, it is possible to have $2\pi f_a<H_I$ which means that the ALPs symmetry breaking takes place after inflation and there will be no iso-curvature perturbations from ALPs. However observations of Lyman-$\alpha$ disfavor FDM\cite{er}.

ALPs with larger mass range ($m_a>10^{-19}$ eV) can be probed from  the observation of the gravitational wave signals from binary merger events at the LIGO-Virgo detectors. For this, detailed analysis of gravitational wave-form and phase are required which will take into account the energy loss by axionic emission.

\section*{Acknowledgments}
The authors thank Kent Yagi for suggesting a correction in the treatment of NS-WD binaries in this paper.

\end{document}